# Fe-Sn nanocrystalline films for flexible magnetic sensors with high thermal stability

Y. Satake[1], K. Fujiwara[1*], J. Shiogai[1], T. Seki[1,2] and A. Tsukazaki[1,2]

[1]*Institute for Materials Research, Tohoku University, Sendai 980-8577, Japan*

[2]*Center for Spintronics Research Network (CSRN), Tohoku University, Sendai 980-8577, Japan*

*Author to whom correspondence should be addressed.
Electronic mail: kfujiwara@imr.tohoku.ac.jp

**Abstract**

The interplay of magnetism and spin-orbit coupling on an Fe kagome lattice in $Fe_3Sn_2$ crystal produces a unique band structure leading to an order of magnitude larger anomalous Hall effect than in conventional ferromagnetic metals. In this work, we demonstrate that Fe-Sn nanocrystalline films also exhibit a large anomalous Hall effect, being applicable to magnetic sensors that satisfy both high sensitivity and thermal stability. In the films prepared by a co-sputtering technique at room temperature, the partial development of crystalline lattice order appears as nanocrystals of Fe-Sn kagome layer. The tangent Hall angle, the ratio of Hall resistivity to longitudinal resistivity, is largely enhanced in the optimal alloy composition of close to $Fe_3Sn_2$, exemplifying the kagome origin even though the films are composed of nanocrystal and amorphous-like domains. These ferromagnetic Fe-Sn films possess great advantages as a Hall sensor over semiconductors in thermal stability owing to the weak temperature dependence of the anomalous Hall responses. Moreover, the room-temperature fabrication enables us to develop a mechanically flexible Hall sensor on an organic substrate. These demonstrations manifest the potential of kagome metal as an untapped reservoir for designing new functional devices.

**Introduction**

Iron-based alloys and compounds have constituted the outstanding basis for applications, particularly with judicious utilization of their rich magnetism and magneto-transport characteristics[1–4]. To enrich their functionality further, extensive investigations have continued on iron-based ferromagnetic materials including ordered alloys[5], oxides[6] and nitrides[7]. In this study, we exemplify magnetic sensor functions in a ferromagnetic iron-tin alloy that is fabricated to harness massive Dirac bands of kagome metal $Fe_3Sn_2$ (ref. [8]). Magnetic sensors are capable of electrically detecting a magnetic field[9] and are becoming increasingly important towards the acceleration of Internet of Things. Their applications include monitoring of electric current via the Oersted field, electronic compasses and motion detection of mechanical parts in microdevices. In conventional semiconductor Hall devices, the detection of a magnetic field (termed instead of magnetic induction, hereafter) $B$ relies on the ordinary Hall effect, which converts a flow of electric current to the transverse Hall voltage $V_{yx}$ (ref. [10]). Since the output $V_{yx}$ is proportional to $B$ and the injection current, good sensor performance under a constant input voltage is achieved in III-V semiconductors such as GaAs, InAs and InSb with high carrier mobility[10]. These semiconductor devices are constructed essentially on highly crystalline films with a low carrier density precisely tuned by high-temperature growth. Their bandgaps are, however, inevitably accompanied by substantial temperature ($T$) dependences of device characteristics. To ensure the stable operation in a wide $T$ range, an external circuit that compensates the $T$ dependence needs to be implemented.

Without changing the basic device structure, the semiconductor can be replaced by a ferromagnetic material with a sufficiently large anomalous Hall effect (AHE). $V_{yx}$ induced by AHE is a nominal function of magnetization $M$, and materials design is better guided with tangent Hall angle, the ratio of Hall resistivity to longitudinal resistivity, rather than mobility and carrier density. In the past two decades, significant progress has been made in the classification of AHEs and understanding of their physical origins[11]. Of particular interest is the intrinsic AHE where Berry curvature arising from electronic band topology acts as an effective magnetic field and can produce a large $V_{yx}$. In this context, a ferromagnetic Fe-Sn alloy, $Fe_3Sn_2$ (Curie temperature $T_C$ = 657 K), is attracting attention because of its very large AHE at room temperature[8,12,13]. The crystal structure consists of alternate stacking of stanene and bilayer of $Fe_3Sn_2$ with a kagome network of Fe, illustrated in Fig. 1a. Recent angle-resolved photoemission spectroscopic study proposed that an interplay of the kagome lattice, which in analogy to graphene produces linearly dispersed bands and Dirac points, and spin-orbit coupling yields massive Dirac bands that concentrate Berry curvature[8]. By positioning the Fermi level within the gap, e.g. with electrostatic gating and impurity doping, quantized AHE[14,15] may be realized at room temperature. $Fe_3Sn_2$ is however thermoequilibrium stable only at temperature above 607 °C; the bulk crystal is formed by a quenching technique[8,12,13,16,17]. Recognizing the uniqueness of $Fe_3Sn_2$, we focus on the thin film of Fe-Sn kagome compounds[18,19] as a candidate for AHE-type Hall devices (Fig. 1b). In this work, we demonstrate magnetic sensor functions of nanocrystalline Fe-Sn alloy films prepared by room-temperature sputtering. Despite the lack of macroscopic lattice order, these films clearly bear characteristics of the crystalline Fe-Sn phase diagram and exhibit AHE as large as those in bulk $Fe_3Sn_2$ kagome metal. This metal-based Hall device can outperform conventional semiconductor Hall devices in thermal stability. The integration on a bendable polymer sheet thanks to the room-temperature fabrication is demonstrated for potential use in flexible electronics. These findings should accelerate challenges to exploitation of exotic physics hosted by iron and other transition-metal kagome compounds[20–23].

**Results**

**Fe-Sn nanocrystalline films grown by co-sputtering at room temperature.** $Fe_xSn_{1-x}$ alloy films were fabricated by a co-sputtering technique.[18,19] The two elements were supplied from a single magnetron cathode by mounting Fe chips on a Sn target (Fig. 1c inset and also see Methods section). In the Fe-Sn binary members[16], antiferromagnetic FeSn (Neel temperature $T_N$ = 368 K) and ferromagnetic $Fe_3Sn$ ($T_C$ = 743 K) also exist, in which the $Fe_3Sn$ kagome layers are accommodated with different stacking sequences (Fig. 1a). To investigate these kagome compounds widely, Fe content $x$ in the films was controlled by the Fe chip number as displayed in Fig. 1c. In-plane distribution of $x$ was as small as a few atomic percent. Figure 1d shows X-ray diffraction (XRD) patterns of films with various $x$, grown on single-crystalline sapphire (0001) substrates at a growth temperature $T_g$ of 500 °C. Nearly equimolar mixture of Fe and Sn ($x$ = 0.54) yields a single phase of FeSn (see Fig. S1 in the Supplementary Information for peak identification). For $x$ = 0.62, diffuse reflections of $Fe_3Sn_2$ are recognized though FeSn still remains. The FeSn phase is also observed even for $x$ = 0.78 whereas the major phase turned to $Fe_3Sn$. Of the three kagome metals, FeSn is thermodynamically stable and the other two are easily decomposed into FeSn and Sn-rich $\alpha$-Fe on cooling after high-temperature crystallization[16]; the above XRD results are consistent with those bulk behavior. As shown in Fig. 1e, the formation of FeSn is found to be suppressed by room-temperature sputtering. Also, macroscopically, the room-temperature grown films do not have crystalline character. Cross-sectional transmission electron microscopy, however, reveals the presence of nanocrystalline domains with typical sizes of as small as a few nanometers, displayed in Fig. 1f (scale bar, 5 nm). Although *d*-spacing values calculated from the selected-area electron diffraction pattern (Fig. 1f inset) are not uniquely indexed with one of the three compounds, the clearly visible layered lattice, combined with the detailed characterization (Fig. S2 in the Supplementary Information), suggests the existence of $Fe_3Sn_2$-like domains in the nanocrystalline film with $x = 0.60$.

**Comparison of anomalous Hall responses for the nanocrystalline film and crystalline film.** Contrary to a naive expectation that the large AHE driven by band topology of the kagome lattice should smear out in such nanocrystalline films, we observed a clear AHE in the room-temperature grown nanocrystalline $Fe_xSn_{1-x}$. Figure 2a and 2b show Hall resistivity $\rho_{yx}$ and magnetization $M$ plots, respectively, measured at 300 K under out-of-plane $B$ application. It is obvious that $\rho_{yx}$ mirrors $M$—AHE is mainly responsible for $\rho_{yx}$. The linear $\rho_{yx}$ response at low magnetic fields and virtually closed hysteresis loop reflect the magnetization vector rotation from the in-plane easy axis to out-of-plane hard axis (Fig. S3 in the Supplementary Information). Notably, the saturation $\rho_{yx}$ does not scale with $x$ dependence of the saturation magnetization. To get insights into the composition dependence, we performed control experiments using high-temperature grown polycrystalline $Fe_xSn_{1-x}$ films ($T_g = 500$ °C), shown in Fig. 2c and 2d. In the polycrystalline films with $x = 0.62$ and $0.78$, the overall anomalous Hall response and saturation magnetic field are similar to those of the nanocrystalline ones. In contrast, AHE and $M$ are considerably small for $x = 0.54$, which is in parallel to the formation of antiferromagnetic FeSn as revealed by XRD. In Fig. 2e, these $x$-dependent AHE properties are summarized using a signature of Hall angle $\rho_{yx} / \rho_{xx}$ ($\rho_{xx}$: longitudinal resistivity). As a consequence of a sharp $\rho_{yx}$ peak around $x = 0.60$ and $\rho_{xx}$ slightly changing with $x$ (Fig. 2e inset), $\rho_{yx} / \rho_{xx}$ takes a broad maximum around $x = 0.60 - 0.75$ in the nanocrystalline films. A weaker but similar trend is also seen for the polycrystalline films that partly contain FeSn, $Fe_3Sn_2$ and $Fe_3Sn$ (Fig. 1d). Note that such a composition dependence is not expected for mere Sn-rich $\alpha$-Fe.

In Fig. 2f, $T$ dependence of Hall conductivity $\sigma_{xy} = \rho_{yx} / (\rho_{xx}^2 + \rho_{yx}^2)$ is compared for nanocrystalline and polycrystalline films with $x \sim 0.6$ and also bulk $Fe_3Sn_2$ in literature[8]. The occurrence of nearly $T$-independent $\sigma_{xy}$ in the polycrystalline films, which resembles the intrinsic behavior in $Fe_3Sn_2$ single crystals[8,13], but with much smaller $\sigma_{xy}$ than the bulk value suggests a small fraction of $Fe_3Sn_2$ crystalized by high-temperature sputtering ($T_g = 500$ °C). The antiferromagnetic FeSn that persistently exists in the polycrystalline film, as found in the decreased $M$ and $\rho_{yx}$ (Fig. 2b, 2d, and also see Fig. S4 in the Supplementary Information), does not give positive contributions to the AHE. In a stark contrast, $\sigma_{xy}$ in the nanocrystalline film rivals the bulk data in the entire $T$ range. One obvious reason for this is the suppression of FeSn by room-temperature sputtering. It is not clear why imperfect development of the kagome lattice order can lead to the large AHE; nevertheless, these observations, together with the $\rho_{yx}$ and $\rho_{yx} / \rho_{xx}$ peaks at $x = 0.60 - 0.75$, strongly suggest that a kagome-derived intrinsic mechanism as proposed for $Fe_3Sn_2$ (ref. [8]) should be the primary origin of the large AHE in nanocrystalline $Fe_xSn_{1-x}$.

**Thermal stability of the AHE for room-temperature deposited films.** Having confirmed the large AHE in nanocrystalline $Fe_xSn_{1-x}$, we now turn to the characterization as a magnetic sensor element. Taking advantages of room-temperature sputtering, we extend our investigation to more commercially available substrates, glass and flexible polyethylene naphthalate (PEN) sheet (Fig. 3a, also see Fig. 1e for their XRD patterns). $T$ dependent AHE characteristics of $Fe_{0.60}Sn_{0.40}$ films with thicknesses $d \sim 40$ nm, displayed in Fig. 3b, are essentially similar on three substrates, demonstrating that specific substrates are not required to achieve the large AHE. Magnetic-field sensing can be performed in the almost linear $\rho_{yx} - B$ region, and the differential coefficient, $\alpha = d\rho_{yx} / dB$ corresponds to the sensitivity for $B$ via $V_{yx}$. As displayed in Fig. 3c and inset, $\alpha$ is nearly constant to a large $B$ of approximately 0.5 T, and is rather insensitive to $T$ variation (red colored regions in the inset). This is more clearly seen in the upper panel of Fig. 3d, where the thermal stability of $\alpha$ is tracked, defined as $\Delta\alpha = (\alpha(T) - \alpha(T = 300 \text{ K})) / \alpha(T = 300 \text{ K})$. In a general operation range of $T = 200 - 400$ K, $\Delta\alpha$ is within a few percent, corresponding to approximately 0.02% / K. The small variation in $\rho_{xx}$, shown in the lower panel, gives an advantage over semiconductor devices that are restricted by inherent thermally activated transport[24].

**Characterization of Hall sensor responses and flexibility of the devices.** To enhance $V_{yx}$ further in view of the film thickness $d$ ($V_{yx} = I \times \rho_{yx} / d$), we examined the lower bound of $d$. Judging from the $d$ dependences of $\rho_{xx}$, $\rho_{yx}$ and $\rho_{yx} / \rho_{xx}$ at $B = 2$ T (Fig. 4a) and also $V_{yx}$ versus $B$ curves (Fig. 4b), we determine that the applicable large AHE persists down to $d = 4$ nm. The $d$ decrease to 2 nm is possible for $\rho_{yx}$, but is accompanied by a sharp rise in $\rho_{xx}$ and the drop of $\rho_{yx} / \rho_{xx}$. These increase power consumption when supplying a constant injection current ($I$). In other words, at the fixed input voltage, $V_{yx}$ is reduced due to the decreased $I$. The 1-nm-thick device was no longer conductive; $d$ of approximately 2 nm may be the critical thickness where island-like domains start to coalesce and form conduction paths. As presented in Fig. 4c, by injecting $I = 10$ mA into the 4-nm-thick device, a large $V_{yx}$ exceeding 0.1 V is generated from a magnetic field of $B = 0.5$ T.

We would like to here note some specific features, which are potentially utilized for three-dimensional magnetic-field sensing. In Fig. 4d, out-of-plane magnetic field angle dependences of $R_{yx}$ under various $B$ are shown. At $B = 9$ T, $R_{yx}$ obeys a $\cos\theta$ relation (black dotted curve) as expected from $M_{\text{eff}} \propto B_{\text{eff}} = B\cos\theta$ with $M_{\text{eff}}$ and $B_{\text{eff}}$ being the out-of-plane components of magnetization and magnetic field, respectively. As $B$ is decreased, because of the in-plane magnetic easy axis (see Fig. S3 in the Supplemetary Information), the actual direction of $M$ vector becomes not to fully follow that of $B$ vector, resulting in a deviation from the $\cos\theta$ relation. Also, in-plane magnetoresistance $R_{\text{sheet}}$ vs $\phi$ shown in Fig. 4e indicates an anisotropic magnetoresistance effect even at low $B = 0.25$ T. By combining these anisotropic responses of $R_{yx}$ and $R_{\text{sheet}}$, the magnetic field vector $\boldsymbol{B}(\theta, \phi)$ could be detected with a simple Hall-bar device.

Nanocrystalline $\text{Fe}_x\text{Sn}_{1-x}$ as demonstrated above can be served as a Hall device-type magnetic sensor. In particular, the capability of sensor integration onto a flexible substrate is appealing, potentially finding applications in flexible electronics[25]. We examined the mechanical bending effect on the nanocrystalline $\text{Fe}_x\text{Sn}_{1-x}$ device. Figure 4f demonstrates that, even under severe bending conditions (see Fig. 4g for the definition of bending geometries), the nanocrystalline $\text{Fe}_{0.60}\text{Sn}_{0.40}$ on PEN offers a reversible operation with an almost unchanged sensor performance (Fig. S5 in the Supplementary Information). Such new functionality enabled by nanocrystalline $\text{Fe}_x\text{Sn}_{1-x}$, in combination with its economically and environmentally friendly ingredients, would offer a new type of magnetic sensor design utilizing AHE.

**Discussion and Conclusions**

The large $\rho_{yx}$ at room temperature has also been obtained in ferromagnetic semiconductors[26] and metal-insulator composites[27]. According to the established classification on the AHE origins[11,28,29], those highly resistive materials, however, are in the poorly conductive region ($\sigma_{xx} < \sim 3 \times 10^3$ $\Omega^{-1}$cm$^{-1}$). Our nanocrystalline Fe$_x$Sn$_{1-x}$ is essentially metal with $\sigma_{xx}$ as high as mid - $10^3$–$10^4$ $\Omega^{-1}$cm$^{-1}$, being in a different category called the intrinsic region (Fig. S6 in the Supplementary Information). In fact, it is observed in nanocrystalline Fe$_x$Sn$_{1-x}$ films that $\sigma_{xy}$ is rather independent of $\sigma_{xx}$, as being consistent with intrinsic mechanisms. At present, the sensitivity of our device is about one order of magnitude lower than those of the state-of-the-art GaAs and Si Hall devices[30]. Based on the Berry curvature scenario, the device performance could be further improved by Fermi-level tuning into the gap at the Dirac point[8]. Such an intrinsic approach, in addition to its critical importance for the next-generation of Hall sensors, may also lead to devices that incorporate exotic quantum transport phenomena, e.g., quantized AHE. We believe that the thin-film structure would be the key enabler for exploration of new functionality that emerges on the kagome lattice.

**Methods**

**Thin-film growth.** Fe$_x$Sn$_{1-x}$ alloy films were fabricated by RF magnetron sputtering. The RF power was set to 50 W, and Ar gas pressure to 0.5 Pa for $x < 0.87$ and 0.8 Pa for $x = 0.87$. A typical growth rate was approximately 4 nm / min as checked by X-ray reflectivity measurement and also with a surface profiler. For films with $d \leq 4$ nm, the surface was covered with a few-nm-thick SiO$_x$ insulating layer to prevent oxidation. The SiO$_x$ layer was formed by RF magnetron sputtering using a SiO$_2$ target at an Ar gas pressure of 0.5

Pa. Composition analysis of the films was performed with energy-dispersive X-ray spectroscopy and inductively coupled plasma atomic emission spectroscopy.

**AHE and magnetization measurements**. Electrical transport properties were measured with a VersaLab, a Physical Property Measurement System (Quantum Design) and a source-measure unit. Films were patterned into a Hall bar structure (1b), and electrical contacts were made with an indium solder. The aspect ratio of electrode-electrode distance for $V_{xx}$ versus that for $V_{xx}$ was approximately unity. To remove thermoelectric and geometric effects, the measured data were symmetrized for $V_{xx}$ and anti-symmetrized for $V_{yx}$ against $B$ as widely adopted to these measurements. Magnetization measurements were carried out using a vibrating sample magnetometry mode of VersaLab.

**Acknowledgements**

The authors thank K. Nakahara, M. Kawasaki, H. Kato, N. Shibata and H. Nishikawa for their helpful advice, and K. Takanashi, S. Ito and F. Sakamoto for their assistance with experiments. This work was partly supported by JSPS KAKENHI (Grants No. 25000003 and JP15H05853) from the Japan Society for the Promotion of Science and Kumagai Foundation for Science and Technology.

**Author contributors**

A.T. and K.F. designed the experiments. K.F. and Y.S. fabricated samples. Y.S., K.F. and J.S. performed electrical measurements. Y.S., J.S. and T.S. contributed to magnetization measurements. Y.S., K.F. and A.T. wrote the manuscript. All authors discussed the results.

**Competing interests**

The authors declare no competing interests.

**Data Availability**

The data that support the findings of this study are available from the corresponding author upon reasonable request.

**Figure legends**

**Figure 1. Sputtered Fe-Sn alloy films. a**, Layered Fe-Sn alloy compounds. The $Fe_3Sn$ kagome layer is depicted in the upper side, and layer stacking in antiferromagnetic FeSn, ferromagnetic $Fe_3Sn_2$ and $Fe_3Sn$ are displayed in the lower side. **b**, The device structure is shown schematically. A excitation current $I$ was injected to an Fe-Sn film on an insulating substrate, and longitudinal voltage $V_{xx}$ and transverse Hal voltage $V_{yx}$ were measured. A magnetic field $B$ is applied perpendicularly to the film plane. **c**, Fe content $x$ in $Fe_xSn_{1-x}$ films was controlled by changing the Fe chip configuration on the Sn target. A photograph when six Fe chips are placed is shown in the inset. Scale bar, 10 mm. Blue, green, and red broken lines correspond to $x = 0.50$ (FeSn), $0.60$ ($Fe_3Sn_2$) and $0.75$ ($Fe_3Sn$), respectively. **d**, XRD patterns for $Fe_xSn_{1-x}$ films with $x = 0.54$, $0.62$ and $0.78$ grown on sapphire (0001) substrates at $T_g = 500$ °C. The data are shifted vertically for clarity. The film thicknesses were approximately 40 nm. See the text and Supplementary Fig. S1 for the phase identification of these films. **e**, XRD patterns for 40-nm-thick $Fe_{0.60}Sn_{0.40}$ films on sapphire, glass and PEN sheet substrates prepared at room temperature. **f,** (false color image) Cross-sectional high-resolution transmission electron microscopy image of a room-temperature sputtered $Fe_{0.60}Sn_{0.40}$ film on a sapphire substrate. The scale bar shows 5 nm. The inset shows a selected area electron diffraction pattern, revealing the presence of nanocrystalline domains.

**Figure 2. AHE in nanocrystalline Fe-Sn alloy films. a,b** Hall resistivity $\rho_{yx}$ (**a**) and out-of-plane magnetization $M$ (**b**) at $T = 300$ K measured for room-temperature sputtered nanocrystalline $Fe_xSn_{1-x}$ films as a function of an out-of-plane magnetic field $B$. The magnetization of a film was extracted by subtracting diamagnetic contributions from the data measured. **c,d** The results for polycrystalline $Fe_xSn_{1-x}$ films grown at $T_g = 500$ °C. **e,** $x$ dependence of Hall angle, $\rho_{yx} / \rho_{xx}$, at $B = 2$ T and $T = 300$ K for nanocrystalline (filled black circles) and polycrystalline $Fe_xSn_{1-x}$ films (open black circles). The inset shows $\rho_{xx}$ (triangles) and $\rho_{yx}$ (squares) for nanocrystalline $Fe_xSn_{1-x}$. **f,** $T$ dependence of Hall conductivity $\sigma_{xy}$ for a nanocrystalline $Fe_{0.60}Sn_{0.40}$ film (filled green circles) and a polycrystalline $Fe_{0.62}Sn_{0.38}$ film (open green circles). For reference, the data of $Fe_3Sn_2$ bulk in literature (ref. [8]) is also included (filled black squares).

**Figure 3. AHE on various substrates. a,** Photograph of a 40-nm-thick nanocrystalline $Fe_{0.60}Sn_{0.40}$ film on a flexible PEN sheet substrate. **b,** $\rho_{yx}$ of 40-nm-thick nanocrystalline $Fe_{0.60}Sn_{0.40}$ films on PEN sheet (magenta), glass (cyan) and sapphire (black) substrates. The AHE measurement was performed at $T = 50, 100, 150, 200, 250, 300, 350$ and $400$ K. The data at 300 K are highlighted with bold lines. **c,** Differential coefficient $\alpha = d\rho_{yx} / dB$. The insets display contour plots of $\alpha$ against $T$ and $B$. **d,** The $T$ variation of $\alpha$ defined as $\Delta\alpha = (\alpha(T) - \alpha(T = 300\text{ K})) / \alpha(T = 300\text{ K})$ and $\rho_{xx}$ are shown in the upper and lower panels, respectively.

**Figure 4. Magnetic sensor properties of nanocrystalline Fe$_{0.60}$Sn$_{0.40}$ films. a,** Thickness (*d*) dependences of $\rho_{xx}$, $\rho_{yx}$ and $\rho_{yx}/\rho_{xx}$ at *B* = 2 T. **b,** *V$_{yx}$* versus *B* curves measured at *I* = 0.1 mA for *d* = 40 nm (black), 20 nm (blue), 10 nm (green), 4 nm (red) and 2 nm (brown). **c,** *V$_{yx}$* output characteristics as a function of *I*. **d,** Out-of-plane magnetic field angle (*θ*) dependence of *R$_{yx}$* for a 4-nm-thick nanocrystalline Fe$_{0.60}$Sn$_{0.40}$ film. The measurement set-up is shown schematically in the inset. The black dotted curve represents a relation $R_{yx} \propto \cos\theta$. **e,** Anisotropic magnetoresistance measurement. Sheet resistance *R$_{sheet}$* was measured in an in-plane *B*. The in-plane rotation angle (*ϕ*) is defined in the inset. The dotted curve is a fitting result using a $\cos 2\phi$ function. **f,** Bending effects on transport properties of 4-nm-thick Fe$_{0.60}$Sn$_{0.40}$ on PEN. See Fig. 4**g** for the definition of *x*-bent and *y*-bent. The sample was first measured without bending (flat, black line), and subsequently characterized under *x*-bent (red) and *y*-bent (blue) conditions. After these cycles, the device recovered back to the initial flat state. **g,** Photographs of the *x*-bent and *y*-bent devices. 4-nm-thick films on PEN sheet substrates were mounted on the surface of a 4.9-mm-radius semicircular jig in two different geometries.

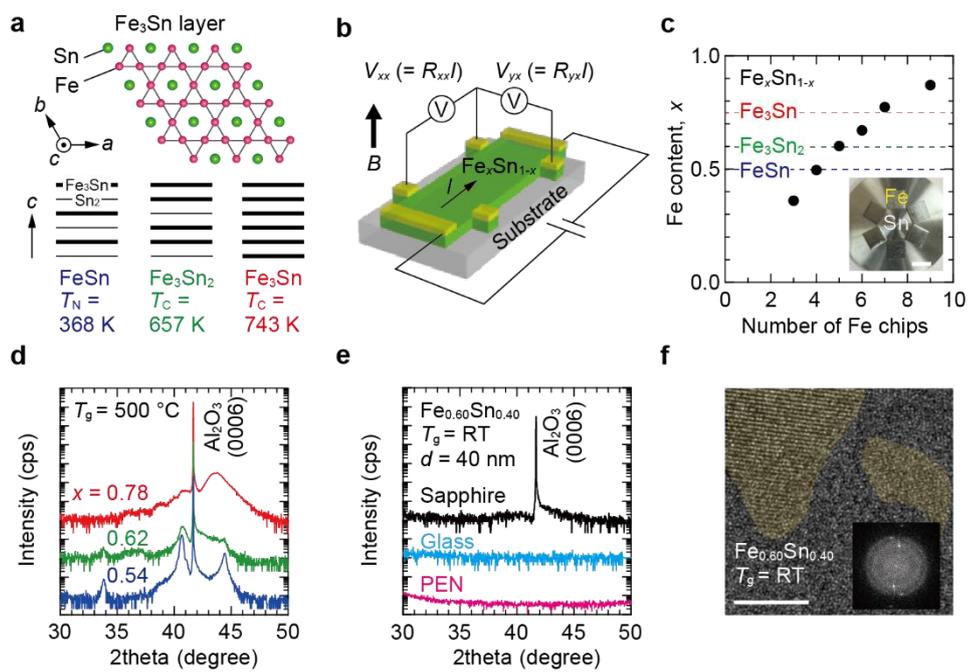

Figure 1. Y. Satake *et al.*

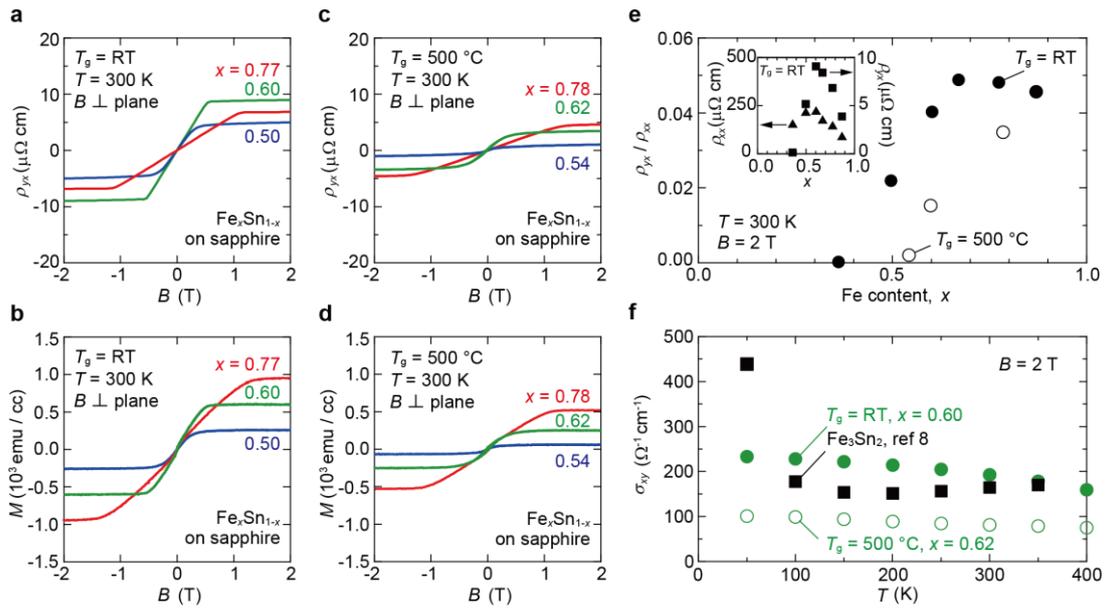

Figure 2. Y. Satake *et al.*

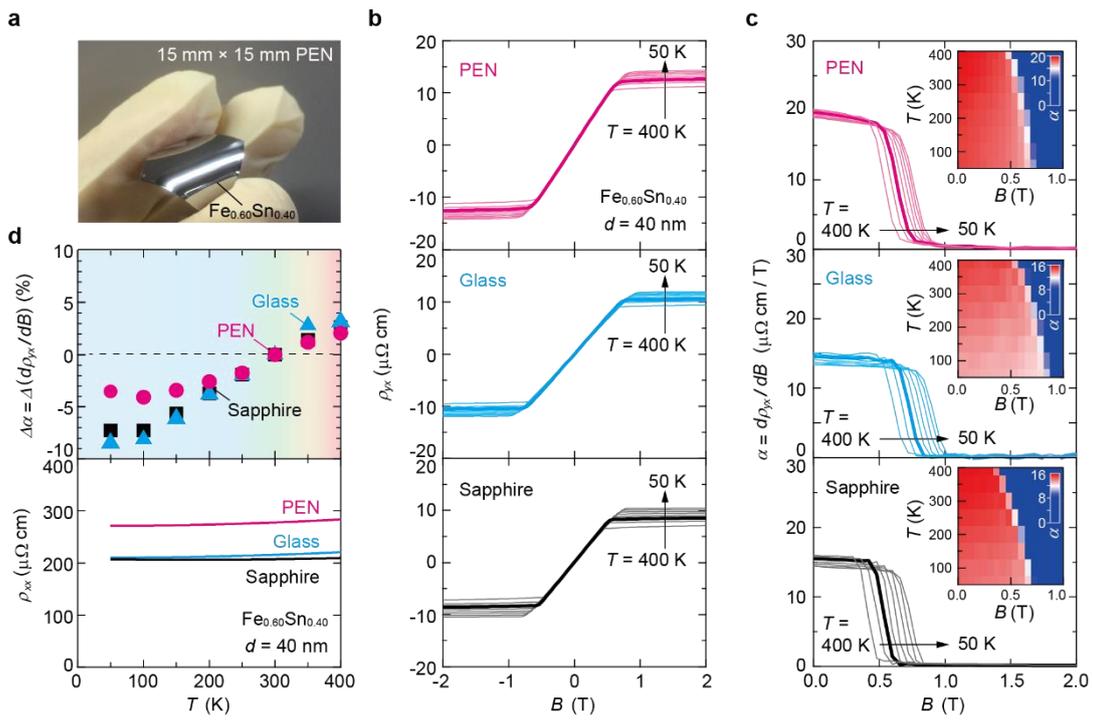

Figure 3. Y. Satake *et al.*

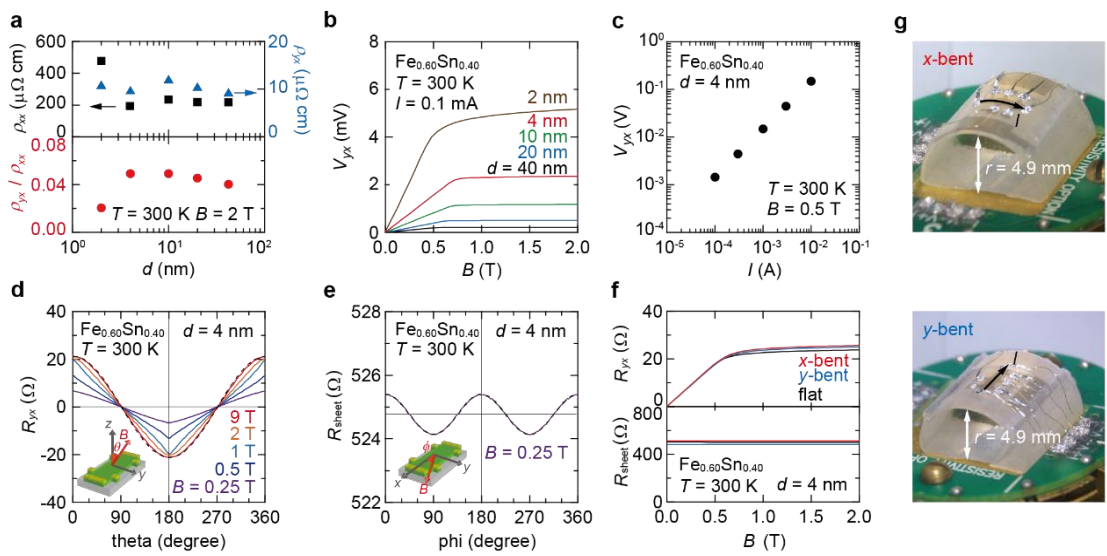

Figure 4. Y. Satake *et al.*